\documentclass[aps,pre,twocolumn,showpacs,longbibliography]{revtex4-1}
\usepackage{color}
\usepackage{amssymb}
\usepackage{amsmath}
\usepackage{bm}
\usepackage{graphicx}

\begin{document}
\title{Conformal invariance of weakly compressible two-dimensional turbulence}

\author{Leonardo Puggioni$^{1}$}

\author{Alexei G. Kritsuk$^{2}$}

\author{Stefano Musacchio$^{1}$}

\author{Guido Boffetta$^{1}$}

\affiliation{$^{1}$Dipartimento di Fisica and INFN, Universit\`a di Torino, 
via P. Giuria 1, 10125 Torino, Italy \\
$^{2}$Center for Astrophysics and Space Sciences, 
University of California San Diego, 9500 Gilman Drive, La Jolla, 
California 92093-0424, USA}

\begin{abstract}
We study conformal invariance of vorticity clusters in weakly compressible
two-dimensional turbulence at low Mach numbers. 
On the basis of very high resolution direct numerical simulation 
we demonstrate the scaling invariance of the inverse cascade with
scaling close to Kolmogorov prediction. 
In this range of scales, the statistics of zero-vorticity isolines 
are found to be compatible with those of critical percolation, 
thus generalizing the results obtained in incompressible 
Navier-Stokes turbulence. 
\end{abstract}


\maketitle 

\section{Introduction}
\label{sec1}
The inverse energy cascade is a key feature of two-dimensional 
incompressible turbulence predicted by Kraichnan many years ago 
\cite{kraichnan1967inertial}.
As a consequence of the existence of two inviscid quadratic invariants 
of the 2D incompressible Navier-Stokes equations,
the enstrophy is transferred to small scales producing the direct cascade,
while energy moves to large scales generating the inverse cascade.
The inverse cascade in 2D incompressible turbulence has been observed 
in numerical simulations 
\cite{herring1985comparison,maltrud1991energy,smith1993bose,
boffetta2000inverse,xiao2009physical}
and laboratory experiments \cite{paret1997experimental,
rivera1998turbulence,chen2006physical},
and its scaling properties have been established including the almost
Gaussian statistics of velocity fluctuations and the absence of intermittency 
\cite{boffetta2000inverse}.

More recently, scaling invariance of the inverse cascade has been promoted
to conformal invariance for some specific features of the turbulent field.
In particular, by using the technique of stochastic-L\"owner evolution
(SLE), it has been shown that clusters of vorticity are statistically
equivalent to those of critical percolation, one of the simplest universality
classes in critical phenomena \cite{bernard2006conformal}.
This result, which suggests an intriguing connection between 
(non-equilibrium) turbulent flows and 
statistical models at the critical point, has been extended to other
2D incompressible flows, including surface quasi-geostrophic turbulence
\cite{bernard2007inverse} and a class of active scalar turbulence 
\cite{falkovich2010conformal}. 
Conformal invariance has also been investigated experimentally in
Lagrangian reconstructed vorticity field of a turbulent surface flow 
\cite{stefanus2011search}
(two-dimensional section of a three-dimensional flow), where 
deviations from SLE predictions have been observed. 

In this paper, we study the appearance of conformal invariance 
in the inverse cascade of weakly compressible 2D turbulence.
Compressible two-dimensional turbulence has applications in numerous
geophysical, astrophysical and industrial problems. Specifically, we
consider a flow with an ideal-gas equation of state with the ratio of
specific heats $\gamma=c_{v}/c_{p} \simeq 1$, which is relevant to 
astrophysical applications (where radiation provides for temperature 
equilibration) \cite{falkovich2017vortices} and for soap films when fluid velocities are of the order of the elastic wave speed in the limit of large
Reynolds numbers \cite{chomaz2001dynamics}.

This paper is organized as follows. In Section~\ref{sec2},
we introduce the physical model, its phenomenology and the numerical 
simulation. In Section~\ref{sec3}, we discuss the statistics of the 
inverse cascade, while Section~\ref{sec4}
is devoted to conformal analysis of isovorticity lines.
Finally, Section~\ref{sec5} contains some conclusions.

\section{Model and phenomenology of 2D compressible flows}
\label{sec2}

The dynamics of a compressible flow is given by
the Euler equations, which impose the
conservation of mass, momentum and total energy:
\begin{equation}
\partial_t \rho + \nabla \cdot \left( \rho {\bm{u}} \right) = 0,
\label{eq:2.1}
\end{equation}
\begin{equation}
  \partial_t \left( \rho {\bm{u}} \right) + \nabla \cdot \left( \rho {\bm{uu}}
  + p \bm I \right) = {\bm{f}},
\label{eq:2.2}
\end{equation}
\begin{equation}
  \partial_t \mathcal{E} + \nabla \cdot \left[ \left(\mathcal{E} +p \right)
    {\bm {u}} \right] = {\bm {f}} \cdot {\bm {u}},
\label{eq:2.3}
\end{equation}
where $\rho$ is the density field,
${\bm {u}}$ the velocity,
$p$ is the pressure,
${\bm {f}}$ is the external forcing,
$\mathcal{E} = \rho \left( u^2/2 +e \right)$ is the total energy density
(the sum of kinetic and potential energy density),
and $\bm{I}$ is the identity matrix.
The system of equations (\ref{eq:2.1}-\ref{eq:2.3})
is closed by the equation of state for an idea gas
$p = \left( \gamma -1 \right) \rho e$.
In the absence of external forcing ${\bm f} = 0$, the system conserves the
total energy $E=\int \mathcal{E} d {\bm {x}}$, which is given by the sum
of the kinetic energy $K=(1/2) \int \rho u^2 d{\bm {x}}$
and the potential energy $U=\int \rho e d{\bm {x}}$.

The average compressibility of the velocity field
is quantified by the rms Mach number $M=\sqrt{\langle u^2 \rangle}/c$,
where $c$ is the speed of sound in the fluid. 
The velocity field can be decomposed
into the solenoidal and irrotational components
${\bm u}={\bm u}_s + {\bm u}_i$,
where ${\bm \nabla} \cdot {\bm u}_s = 0$
and ${\bm \nabla} \times {\bm u}_i = 0$. 

In the case of a two-dimensional flow, 
they can be expressed in terms of two scalar fields: 
the stream function $\psi$, and the velocity potential $\phi$, 
\begin{eqnarray}
{\bm u}_s &=& (\partial_y \psi, -\partial_x \psi), 
\label{eq:3.1} \\
{\bm u}_i &=& {\bm \nabla} \phi\;.   
\label{eq:3.2}
\end{eqnarray}

The Laplacian of the velocity potential provides a local measure of the 
divergence of the velocity ${\bm \nabla} \cdot {\bm u}=\nabla^2 \phi$,
while the Laplacian of the stream functions defines the vorticity field
$\omega = \partial_x u_y - \partial_y u_x = -\nabla^2 \psi$.

The phenomenology of a 2D compressible flow is strongly dependent
on the Mach number.
In the low-compressibility regime ($M \ll 1$), the behavior is similar
to that of the incompressible case.
One observes the development of a double-cascade scenario 
in which the enstrophy
$\Omega = (1/2) \int \omega^2 \rho d{\bm {x}}$
is preferentially transferred toward small scales (direct cascade),
while the kinetic energy is transferred mostly to large scales
(inverse cascade).
In the absence of a large-scale dissipation mechanism
(such as friction),
the inverse cascade process causes the accumulation of energy at
the largest scale (smaller wavenumber) of the flow.
The phenomenon of ``spectral condensation'' of kinetic energy on the
lowest accessible wavenumber has been studied
both in the 2D incompressible~\cite{smith1993bose}
and compressible flows~\cite{falkovich2017vortices,naugolnykh2014nonlinear}.
It has also been observed in quasi-2D geometries, i.e. in the
turbulent dynamics of thin 
layers~\cite{xia2009spectrally,musacchio2019condensate}.

The energy accumulation at large scales causes the growth in time
of the Mach number, increasing the compressibility of the flow.
A peculiar phenomenon of the compressible case is the formation
of acoustic waves, i.e. pressure fluctuations which propagate within
the fluid~\cite{lighthill1955effect}. 
Acoustic waves of sufficiently large amplitude break to form a train of
N-waves. Emerging shocks in turn speed up the attenuation of acoustic energy.
Shocks also amplify small-scale vorticity by compression and produce new
vorticity through shock-shock interactions, while strong shear generates new
shocks and rarefaction waves \cite{miles1957on,artola1989nonlinear}.
At sufficiently large Mach numbers, the interaction between acoustic waves 
and vortices thus causes a transfer of energy toward small-scales 
through wave breaking and
the generation of shocks~\cite{falkovich2017vortices}.
This provides a stabilizing mechanism for the energy of the condensate,
which is fed by the inverse cascade process and
is removed by the acoustic waves,
allowing for the formation of a statistically steady state.
This process resembles the so-called ``flux loop'' observed in
2D stratified flows~\cite{boffetta2011fluxloop}. 
While the dynamics of vortices and waves is strongly coupled at large scales, 
it has been found to be almost independent at small scales,
where the cascade of wave energy follows the predictions 
of acoustic turbulence and is decoupled from
the enstrophy cascade~\cite{falkovich2017vortices}.

\subsection{Numerical simulation}
\label{sec2a}

The Euler equations (\ref{eq:2.1}-\ref{eq:2.3})
have been integrated by an implicit 
large eddy simulation (ILES) \cite{sytine2000convergence} 
in a square periodic domain of size $L$
on a grid of $8192 \times 8192$ points using an
implementation of the piecewise parabolic method (PPM) 
\cite{colella1984piecewise}
with the {\it Enzo} code \cite{bryan2014enzo}.
The reference length, time, and mass in the simulation 
are defined by choosing the box size $L=1$,
the speed of sound $c=1$, and the mean density $\rho_0=1$. 
Starting from a zero initial velocity field, the system is forced 
by a solenoidal, random external forcing ${\bm f}$ acting on am 
intermediate pumping scale $L_f = 2 \pi / k_f$ with $k_f = 1024 \pi$.
The time correlation of the forcing is of the order of the 
time step, i.e. much smaller to any physical timescale in the system.
The rate of kinetic energy injection provided by the forcing is
$\varepsilon_f = 0.001$.
We remark that this value has to be kept
sufficiently small to avoid the production of shock waves
at the injection scale, which would inhibit the inverse cascade of energy.
The characteristic vortex turn-over time at the scale $L_f$ is
$\tau_f=\rho_0^{1/3} L_f^{2/3}\varepsilon_f^{-1/3 }\simeq 0.52$,
which is more than $10^4$ times larger than the forcing correlation time.
The time integration has been performed up to time $t=30$ with
a sampling of the velocity field and computation of the vorticity
field every $\Delta t = 0.05$.

We remark that even if no explicit dissipative mechanism is prescribed in 
(\ref{eq:2.1}-\ref{eq:2.3}), the code introduces numerical 
dissipation which strongly affects scales smaller than $16 \Delta x$ 
($\Delta x=1/8192$ is the spatial resolution). 

\section{Statistics of the inverse cascade}
\label{sec3}

\begin{figure}[h!]
\includegraphics[width=0.9\columnwidth]{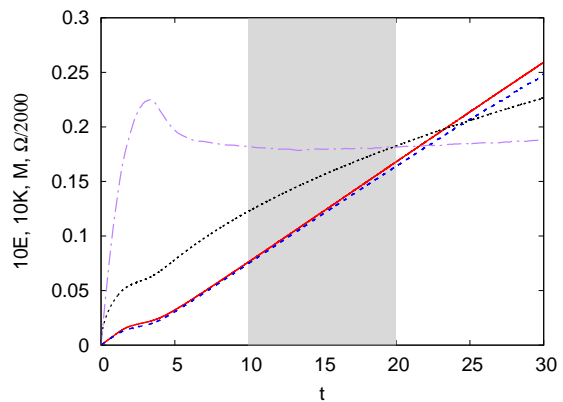}
\caption{Time evolution of the
  total energy $E$ (red solid line),
  kinetic energy $K$ (blue dashed line),
  enstrophy $\Omega$ (purple dash-dotted line),
  and Mach number $M$ (black dotted line). 
  The values of $E$ and $K$ have been multiplied by a factor of 10,
  and $\Omega$ by a factor of $1/2000$ for plotting purposes.
  }
\label{fig:1}
\end{figure}

The temporal evolution of the total energy $E$, the kinetic energy $K$, 
the enstrophy $\Omega$, and the Mach number $M$ during the simulation 
is shown in Figure~\ref{fig:1}. 
At the beginning of the simulation, the Mach number is very small 
and the dynamics of the system are dominated by its incompressible part. 
Therefore, we expect to observe the development of an inverse energy cascade 
propagating from the forcing scale $L_f$ to larger scales $r \gg L_f$ 
and a direct enstrophy cascade toward small scales $r \ll L_f$.  
This is confirmed by the linear growth of the total energy $E \sim \varepsilon_{inv} t$
with a growth rate $\varepsilon_{inv}  \approx 0.92 \varepsilon_f$,
which corresponds to the flux of energy in the inverse cascade. 
The total energy is dominated by the contribution of the kinetic energy $K$, 
while the potential energy $U$ becomes visible only at late times $t \gg 20$. 

In contrast to the energy, we find that after an initial growth, 
the enstrophy $\Omega$ reaches an almost constant value (see Fig.~\ref{fig:1}). 
This happens because the direct enstrophy cascade transfers
the enstrophy injected by the forcing to the small scales,  
where it is removed by numerical dissipation.

\begin{figure}[h!]
\includegraphics[width=0.9\columnwidth]{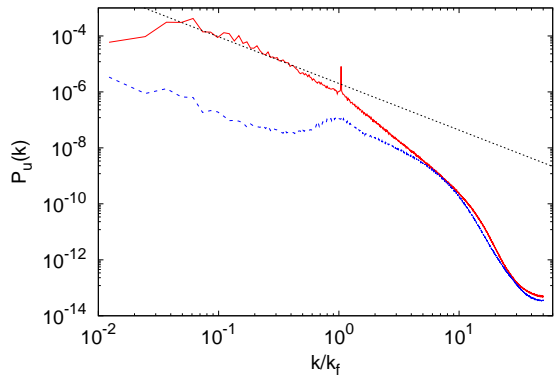}
\caption{
Power spectra of the solenoidal component of the velocity $P_s(k)$ 
(red solid line) and of the irrotational component $P_i(k)$
(blue dashed line), at time $t = 20$. Black dotted line represents
Kolmogorov scaling $k^{-5/3}$.
}
\label{fig:2}
\end{figure}

At late times,  the inverse energy cascade will eventually produce 
an accumulation of the energy at the largest scale, giving rise to 
the formation of an intense vortex dipole (the condensate). 
The formation of this vortical structure results in non-Gaussian 
statistics in the vorticity field and causes the breaking of scale 
invariance~\cite{kritsuk2019energy}.
Considering that we are interested in the study of the conformal invariance 
of the vorticity field, it is crucial to verify that the field is at 
least scale-invariant. 
For this reason, in the following we will limit our analysis to the 
time range $10 < t < 20$, 
i.e., well before the beginning of the formation of the condensate. 
Moreover, in this time interval the value of $\Omega$ is almost constant, 
which allows us to assume that the dynamics of the vorticity field are in 
a statistically steady state. 
In this time range, the Mach number varies from $M \simeq 0.12$ at $t=10$ to
$M \simeq 0.18$ at $t=20$ (Fig.~\ref{fig:1}). 
The dynamics are therefore weakly compressible. 

Furthermore, we will consider the scaling properties in the range of scales 
of the inverse cascade $L_f \ll r \ll L$, 
in which the conformal invariance has been detected 
for the case of 2D incompressible turbulence~\cite{bernard2006conformal}. 
As shown in Figure~\ref{fig:2}, 
in the range of wavenumbers of the inverse cascade ($0.04k_f<k<k_f$) 
the power spectrum of kinetic energy is dominated 
by the spectrum of the solenoidal component of the velocity field 
$P_s(k)   =   \sum_{|q|=k} |{\bm u}_s(q)|^2 $, 
while the spectrum of the irrotational component 
$P_i(k) =  \sum_{|q|=k} |{\bm u}_i(q)|^2 $ is much smaller  
(by more than a factor of $200$ between 
$0.04k_f<k<0.4k_f$).
The power spectrum of the solenoidal component $P_s(k)$ 
displays an approximatively Kolmogorov slope for wavenumbers 
$k<0.6k_f$
with a steeper exponent close to $-2$ close to the forcing
scale \cite{falkovich2017vortices,kritsuk2019energy}. 
At high wavenumbers, the spectra of the irrotational and 
solenoidal components become comparable. 
With a more accurate numerical method and adaptively controlled numerical
dissipation, two independent direct cascades can be resolved at $k>k_f$: the
enstrophy cascade and the acoustic energy cascade. These are reflected in the
scaling of power spectra, $P_s(k)\sim k^{-3}\ln{(k/k_f)}$ and $P_i(k) \sim
k^{-2}$ \cite{kritsuk2019energy}. While $P_s(k)$ dominates at $k\lesssim k_f$,
$P_i(k)$ inevitably becomes dominant at $k\gg k_f$.

\begin{figure}[h!]
\includegraphics[width=0.9\columnwidth]{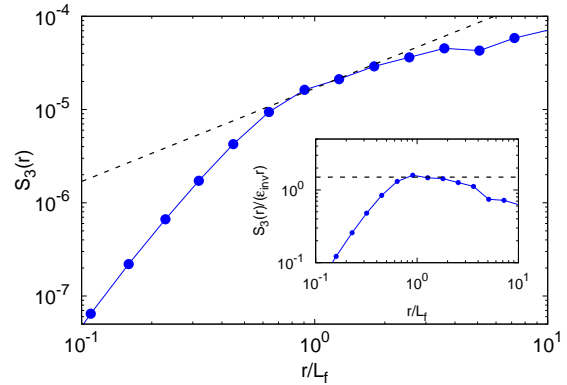}
\caption{  
Third-order longitudinal structure function $S_3(r)$ (blue circles)
average over times $10 \le t \le 20$. Black dashed line represents
(\ref{eq:3.3}).
Inset: compensated structure function
$S_3(r)/(\varepsilon_{inv} r)$ (blue circles)
and the predicted value $3/2$ (black dashed line).   
}
\label{fig:3}
\end{figure}

Another indication of the scale invariance of the velocity field in the range
of scales of the inverse energy cascade is provided by the third-order 
longitudinal structure function (SF)
$S_{3}(r) = \langle (\delta u_L (r) )^3 \rangle$ where 
$\delta u_L(r)=[{\bm u}({\bm x}+{\bm r})-{\bm u}({\bm x})]\cdot {\bm r}/r$
is the longitudinal velocity difference at scale $r$.
In 2D incompressible turbulence, the constant energy flux
in the inertial range gives an exact prediction for the third-order
structure function which, in homogeneous and isotropic conditions reads
\cite{bernard1999three,lindborg1999can}
\begin{equation}
S_3(r) = {3 \over 2} \varepsilon_{inv} r,
\label{eq:3.3}
\end{equation}
where $\varepsilon_{inv}$ represent the inverse kinetic energy flux.
The third-order SF, time averaged for $10 < t < 20$ in our simulation, 
is shown in Figure~\ref{fig:3}. 
It displays a linear scaling range at $r > L_f$ 
with a coefficient ${3 \over 2} \varepsilon_{inv}$ (see inset of Fig.~\ref{fig:3})
in agreement with the assumption of a constant energy flux. 
Let us notice that, because of the lack of stationarity at large 
scales $r \gg L_f$, the scaling $S_3(r) \sim r$ is observed only in 
a narrow range of scales. 
  
\section{Conformal invariance of isovorticity lines}
\label{sec4}
The discovery of conformal invariance in two-dimensional turbulence was
 first made for the zero-vorticity lines in incompressible Navier-Stokes
equations \cite{bernard2006conformal} and then extended to others
two-dimensional turbulent systems characterized by different scaling
laws \cite{bernard2007inverse,falkovich2010conformal}.
These previous results suggest the possibility to also test conformal 
invariance in the inverse cascade of weakly compressible turbulence.

\begin{figure}[h!]
\includegraphics[width=0.9\columnwidth]{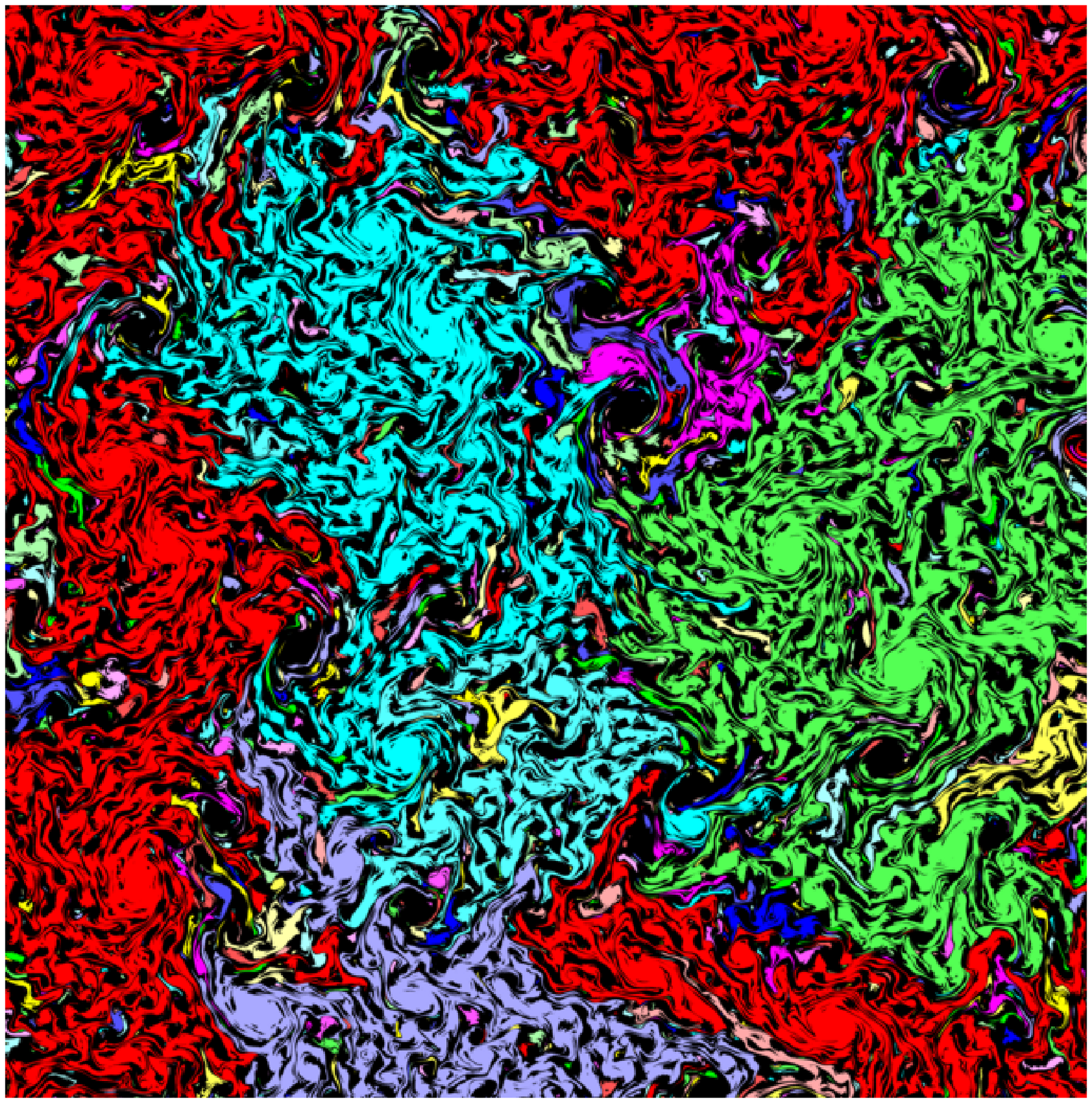}
\includegraphics[width=0.9\columnwidth]{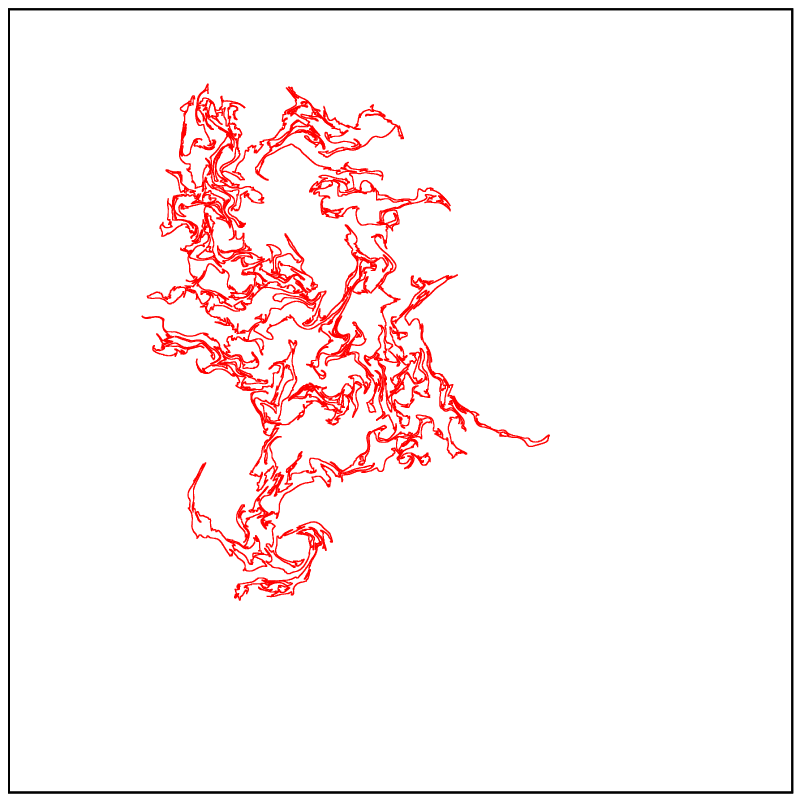}
\caption{
Top:
Vorticity cluster, defined as connected regions with the 
same sign of vorticity (here positive). Different colors 
are attributed to different clusters. Regions of negative 
vorticity are black. 
Bottom:
Zero-vorticity isoline of the cyan cluster.
}
\label{fig:4}
\end{figure}
We have extracted the vorticity clusters (i.e. connected regions
of positive/negative vorticity) and zero vorticity isolines (boundaries 
of vorticity clusters) from the different fields of the simulation.
We have obtained an ensemble of $N_c=461,399$ clusters. 
One example of these clusters is shown in Fig.~\ref{fig:4} for an intermediate
time in the simulation $t=20$. 
Here, we observe the presence of clusters of 
different sizes, each one enclosed by a complex, fractal boundary.

\begin{figure}[h!]
\includegraphics[width=0.9\columnwidth]{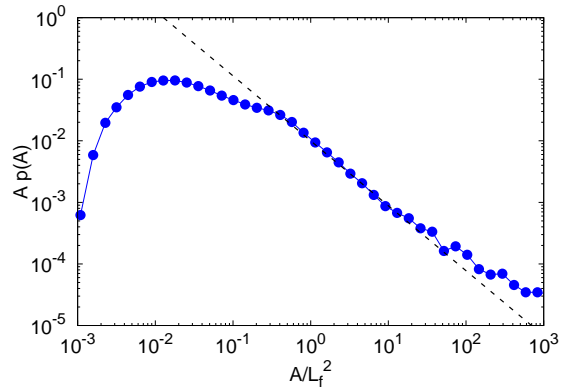}
\caption{
Probability density function of cluster size $A$ (blue circles). 
The black dashed line represents the scaling prediction 
by percolation theory
$A p(A) \sim A^{-96/91}$.  
}
\label{fig:5}
\end{figure}

Figure~\ref{fig:5} shows the probability distribution function (PDF) $p(A)$ 
of cluster size $A$, defined as the number of connected sites which belong 
to the cluster.  The PDF displays a power-law behavior 
in the range $ L_f^2 < A < 10 L_f^2$. 
The scaling exponent observed in Fig.~\ref{fig:5}  is in agreement with 
the theoretical value predicted in the case of critical percolation in 2D
$p(A) \sim A^{-96/91-1}$.~\cite{stauffer2018introduction} 
The same value has been previously observed for the scaling exponent of the PDF 
of vorticity clusters size in the case of incompressible 
2D Navier-Stokes~\cite{bernard2006conformal}.
 
This result suggests that vorticity clusters produced 
by the inverse cascade in weakly compressible turbulence are statistically 
equivalent to clusters of critical percolation and therefore display
the same properties of conformal invariance. In particular, the cluster
boundaries in the continuous limit are expected to belong to the class of
conformal curves called SLE curves 
\cite{schramm2000scaling,cardy2005sle}.
In order to introduce briefly the basics of the SLE,   
let us consider a curve $\gamma(s)$, parameterized by the time $s$,  
starting from a point of the boundary of the half-plane $H$.
At given time $s$, the curve $\gamma(s)$ define a region $K_s$
(the {\it hull}) formed by the points of the complex half-plane which cannot
be reached from infinity without crossing the curve, plus the curve itself.
The simply connected set $H \setminus K_s$ can be mapped into $H$
by an analytic function $g_s(z)$ which satisfies
the asymptotic behavior
$g_s(z) \sim z+2s/z+O(z^{-2})$ at $z\to \infty$.
The conformal map $g_s(z)$ obeys the differential
L\"owner equation~\cite{lowner1923untersuchungen}:
\begin{equation}
\frac{d g_s}{ds} = \frac{2}{g_s(z)-\xi(s)},
\label{eq:4.1}
\end{equation}
where $\xi(s)$ is the real {\it driving function}.
The L\"owner equation establishes an equivalence
between the curve $\gamma(s)$ and its driving function $\xi(s)$.
Different driving functions produce different curves.
In the case of random curves $\gamma(s)$, the equation~(\ref{eq:4.1})
is called stochastic Loewner evolution (SLE) and the driving
$\xi(s)$ is a random real variable. 
It has been demonstrated that the statistics of random curves 
are conformal invariant if and only if the driving is
a Brownian walk, i.e. a random function with independent increments and 
with $\langle (\xi(s)-\xi(0))^2 \rangle = \kappa s$. Here, $\kappa$ is the
diffusion coefficient which classifies the universality class of cluster 
boundaries in critical phenomena in 2D 
\cite{bauer2003conformal,gruzberg2004loewner,cardy2005sle}. 
One of the predictions for SLE curves is their fractal dimension, which
is known to be $D=1+\kappa/8$ (for $\kappa<8$). In the case of 
critical percolation, for which $\kappa=6$, the prediction is 
$D=7/4$ which has been indeed measured in the vorticity cluster of
two-dimensional turbulence\cite{bernard2006conformal}.

We have therefore extracted the zero-vorticity line from the fields of weakly 
compressible turbulence.
The extraction is performed by means of an algorithm which follows the
frontier of a cluster of vorticity by always keeping
the positive region on the right of the path. 
At variance with ``true'' SLE curves, in our numerical simulation, scale
(and conformal) invariance can be expected in the range of scales of the 
inverse cascade only. Therefore, for numerical convenience, we have
coarse-grained the vorticity fields produced by the simulation by halving
the resolution to a $4096 \times 4096$ grid.
One example of vorticity isoline obtained from this procedure 
is shown in Fig.~\ref{fig:4} (right panel).
The total number of isolines obtained is $N=1144$. 

\begin{figure}[h!]
\includegraphics[width=0.9\columnwidth]{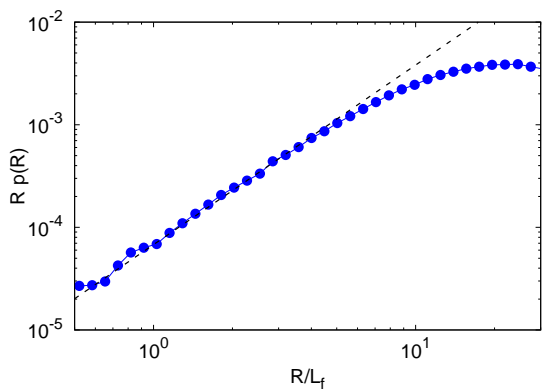}
\caption{Probability density function of distances $R$ between 
two points belonging to the the same zero-vorticity line (blue circles).
The black dashed line represents the theoretical prediction 
$R p(R) \sim R^{7/4}$.
}
\label{fig:6}
\end{figure}

We have computed the correlation dimension $D_2$ of
the zero-vorticity isolines by computing the probability density function (PDF) $p(R)$ 
of finding two points belonging to the same isoline at distance $R$. 
For a fractal set, the probability scales as $p(R) \propto R^{D_2-1}$.
As shown in Fig.~\ref{fig:6}, the PDF $p(R)$
displays a scaling exponent in agreement with the prediction for the
SLE curves with $\kappa =6$, i.e., $D_2 = 7/4$ in the range of scales
$1 \lesssim R/L_f \lesssim 7$, which corresponds to the
the inverse cascade. 

\begin{figure}[h!]
\includegraphics[width=0.9\columnwidth]{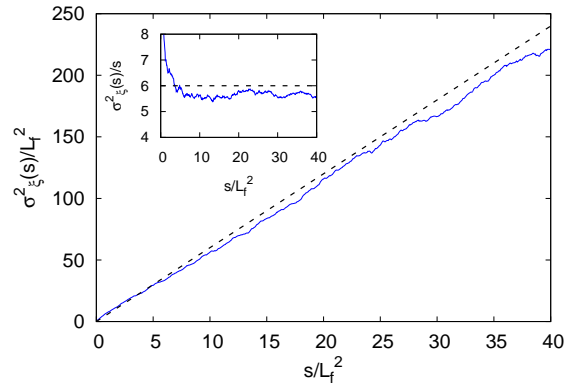}
\caption{Variance of driving function$\sigma^2_\xi(s)$ as a function
  of the parameterization $s$ (blue solid line)
  and the theoretical prediction $\sigma^2_\xi(s) = \kappa s$ with $\kappa=6$
  (black dashed line).
  In the inset we show the compensated value $\sigma^2_\xi(s)/s$. 
}
\label{fig:7}
\end{figure}

\begin{figure}[h!]
\includegraphics[width=0.9\columnwidth]{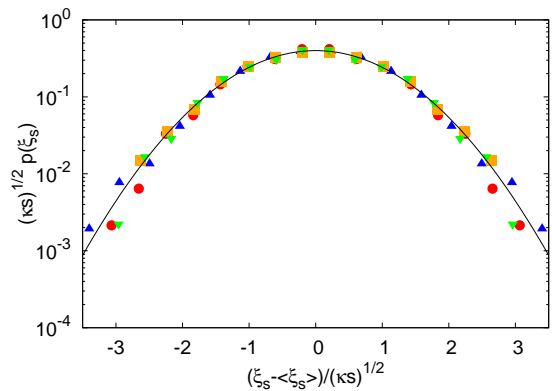}
\caption{PDF of the driving functions $\xi_s$
  at different values of $s$:
  $s = 10 L_f^2$ (blue upper triangles),
  $s = 20 L_f^2$ (red circles),
  $s = 30 L_f^2$ (orange squares),
  $s = 40 L_f^2$ (green lower triangles). 
}
\label{fig:8}
\end{figure}

Assuming that the ensemble of zero-vorticity isolines is statistically
equivalent to SLE curves, we have derived the associated ensemble of
driving functions $\xi(s)$.
The algorithm which computes the driving for a generic curve
is based on the solution of the Eq.~(\ref{eq:4.1})
in the case of an infinitesimal line segment starting
from the origin $(0,0)$ and ending in $(\xi,2\sqrt{ds})$: 
$g_{ds}(z) = \xi + \sqrt{(z-\xi)^2+4ds}$.
By approximating the generic curve $\gamma(s)$ with piecewise
line segments, one obtains the associated driving function $\xi(s)$
(see~\cite{bernard2006conformal} for further details).
Averaging over the ensemble of $N=1144$ driving functions obtained,
we have computed the variance
$\sigma^2_\xi = \langle (\xi(s)-\langle \xi(s) \rangle)^2 \rangle$.
In the range $1 \lesssim s/(2 L_f)^2 \lesssim 7$,
we find that the variance grows linearly as
$\sigma^2_\xi = \kappa s$ (see Fig.\ref{fig:7}).
The driving $\xi(s)$ is therefore a diffusive process
with $\kappa \approx (5.7 \pm 0.2)$,
which is close (within the statistical uncertainty)
to the expected value of $\kappa=6$.
Moreover, the PDF
of the standardized driving
$(\xi(s) - \langle \xi(s) \rangle)/(\kappa s)^{1/2}$ collapses
onto a standard Gaussian distribution function for values of $s$
in the scaling range (see Fig.~\ref{fig:8}).
These findings support the conjecture that the driving function
is a genuine Brownian motion $\xi_s = \sqrt{\kappa}B_s$, 
and that the vorticity isolines are
SLE-curves belonging to the same class of universality of percolation corresponding to 
$\kappa=6$.


\section{Conclusions}
\label{sec5}
In this work we have studied the conformal invariance 
of weakly compressible two-dimensional turbulence. 
We have shown that the isolines 
of vorticity clusters are compatible with SLE curves
in the universality class of critical percolation, as 
in the case of incompressible two-dimensional turbulence.
Our results therefore extend those obtained in others
two-dimensional turbulent systems (Navier-Stokes, surface 
quasi-geostrophic, Charney-Hasegawa-Mima) to the realm of
(weakly) compressible turbulence. 

One question which naturally arises from our results is whether 
the conformal invariance property would survive in higher 
compressible regimes characterized by larger Mach numbers. 
While this is in principle interesting, we remark that, by
increasing the Mach number, a different 
phenomenology emerges at large scales: when velocity fluctuations reach the
speed of sound, kinetic energy produces shock waves which
arrest the inverse cascade and provide a new mechanism of energy 
dissipation \cite{falkovich2017vortices}. 
This ``flux loop'' introduces a characteristic scale in the
process which breaks the scaling invariance of the cascade.
Conformal (and scaling) invariance could still survive in the 
limited range of scales between energy injection and shock
wave production, and its study would be an interesting problem
for future investigation.

\begin{acknowledgments}
G.B. and S.M. acknowledge financial support by the
Departments of Excellence grant (MIUR). A.K. acknowledges
support by the Binational Science Foundation under Grant No. 2018118.
This work used the Extreme Science and Engineering Discovery Environment 
(XSEDE), including the {\em Comet} and {\em Data Oasis} systems at 
the San Diego Supercomputer Center, through allocation TG-MCA07S014
and Director's Discretionary allocation DDP189.
\end{acknowledgments}

\bibliography{biblio}

\end{document}